\documentclass[10pt, letterpaper, notitlepage, oneside]{article}

\usepackage[
	letterpaper,
	bottom      = 1.00in,
	left        = 0.75in,
	right       = 0.75in,
	top         = 1.00in
]{geometry}
\usepackage[utf8]{inputenc}
\usepackage[nopar]{lipsum}

\usepackage{fancyhdr}
\fancypagestyle{specialfooter}{
	\fancyhf{}
	\lhead{Published in \textit{Fluid Dynamics Research}: \href{https://doi.org/10.1088/1873-7005/aa6634}{\url{https://doi.org/10.1088/1873-7005/aa6634}} \hfill \thepage}
	\cfoot{\textbf{Email addresses for correspondence:} $\dag$ g\_dilabb{\MVAt}encs.concordia.ca; $\ddag$ lcfd{\MVAt}encs.concordia.ca}
}

\usepackage{hyperref}
\hypersetup{
	breaklinks = true,
	citecolor  = blue,
	colorlinks = true,
	filecolor  = blue,
	linkcolor  = blue,      
	urlcolor   = blue
}

\usepackage{titlesec}
\titleformat*{\section}{\Large\bfseries}
\titleformat*{\subsection}{\normalsize\bfseries\filcenter}
\titleformat*{\subsubsection}{\normalsize\bfseries}

\makeatletter
\renewcommand\@seccntformat[1]{\csname the#1\endcsname.\quad}
\makeatother

\usepackage{caption}
\usepackage{cleveref}
\usepackage{graphicx}
\usepackage{epstopdf}
\usepackage[
	font = normalsize
]{subfig}
\usepackage{xparse}
\crefformat{subfigure}{#2\onlyletter{#1}#3}
\crefrangeformat{subfigure}{#1--#3\onlyletter{#2}#4}
\ExplSyntaxOn
\NewDocumentCommand{\onlyletter}{m}{
	\tl_set:Nx  \l_tmpa_tl { #1 }
	\tl_item:Nn \l_tmpa_tl { -1 }
}
\ExplSyntaxOff

\usepackage{amsfonts}
\usepackage{amsmath}
\usepackage{marvosym}
\usepackage{textcomp}
\usepackage{upgreek}
\newcommand{\volume}{{\ooalign{\hfil$V$\hfil\cr\kern0.08em--\hfil\cr}}}

\usepackage{arydshln} 
\usepackage{multirow}
\usepackage{setspace}

\usepackage{apacite} 
\let\cite\shortcite
\let\citeA\shortciteA

\newcommand{\noop}[1]{}

\begin{document}
	
	\thispagestyle{specialfooter}
	
	\noindent
	\textbf{\LARGE Numerical simulation of flows in a circular pipe transversely\\\\ subjected to a localized impulsive body force with applica-\\\\ tions to blunt traumatic aortic rupture}\hfill\break
	
	\begin{changemargin}{0.25in}{0.00in}
		{\large Di Labbio G$^{1\dag}$, Keshavarz-Motamed Z$^{1,2}$, Kadem L$^{1\ddag}$}\hfill\break
		$^{1}$\textit{Laboratory of Cardiovascular Fluid Dynamics, Concordia University, Montr\'{e}al, QC, Canada, H3G 1M8}\hfill\break
		$^{2}$\textit{Department of Mechanical Engineering, McMaster University, Hamilton, ON, Canada, L8S 4L8}\hfill\break
		
		Much debate surrounds the mechanisms responsible for the occurrence of blunt traumatic aortic rupture in car accidents, particularly on the role of the inertial body force experienced by the blood due to the abrupt deceleration. The isolated influence of such body forces acting on even simple fluid flows is a fundamental problem in fluid dynamics that has not been thoroughly investigated. This study numerically investigates the fundamental physical problem, where the pulsatile flow in a straight circular pipe is subjected to a transverse body force on a localized volume of fluid. The body force is applied as a brief rectangular impulse in three distinct cases, namely during the accelerating, peak, and decelerating phases of the pulsatile flow. A dimensionless number, termed the degree of influence of the body force ($\Psi$), is devised to quantify the relative strength of the body force over the flow inertia. The impact induces counter-rotating cross-stream vortices at the boundaries of the forced section accompanied by complex secondary flow structures. This secondary flow is found to develop slowest for an impact occurring during an accelerating flow and fastest during a decelerating flow. The peak skewness of the velocity field, however, occurred at successively later times for the three respective cases. After the impact, these secondary flows act to restore the unforced state and such dominant spatial structures are revealed by proper orthogonal decomposition of the velocity field. This work presents a new class of problems that requires further theoretical and experimental investigation.
		\hfill\break
		
		\hfill\break
		* \textit{Data pertaining to this article will be made available from the authors upon reasonable request.}\hfill\break
		* \textit{The authors have no conflicts of interest to declare.}\hfill\break
		* \textit{Please cite as}: Di Labbio, G., Keshavarz-Motamed Z, \& Kadem, L. (2017). Numerical simulation of flows in a circular pipe transversely subjected to a localized impulsive body force with applications to blunt traumatic aortic rupture. \textit{Fluid Dynamics Research}, \textit{49}(3), 035510.
		
		\hfill\break
		\raisebox{0.95pt}{\small\textcopyright} 2017 IOP Publishing. This manuscript version is made available under the CC BY-NC-ND 3.0 license, more information regarding usage terms can be found at \href{https://creativecommons.org/licenses/by-nc-nd/3.0/}{\url{https://creativecommons.org/licenses/by-nc-nd/3.0/}}. The published version of the manuscript is available at \href{https://doi.org/10.1088/1873-7005/aa6634}{\url{https://doi.org/10.1088/1873-7005/aa6634}}. The supplementary information associated with this manuscript can be found at \href{http://stacks.iop.org/FDR/49/035510/mmedia}{\url{http://stacks.iop.org/FDR/49/035510/mmedia}}.
	\end{changemargin}
	
	\section{\label{sec:Intro}Introduction}
		
		Blunt traumatic aortic rupture (BTAR) is the seconding leading cause of death in car accidents in North America, responsible for $7500$ to $8000$ casualties per year \cite{Richens02}. Victims of BTAR have a rather grim fate with an estimated $70$ to $90$\% of patients dying on site, and of those fortunate survivors an additional $40$ to $50$\% are estimated to succumb to their injuries within $24$ h \cite{Chiesa03}. Although vehicle safety systems have undoubtedly advanced over the past decade, their effectiveness in preventing BTAR is minor, as the incidence of BTAR has remained unchanged regardless of airbag system and seat belt use \cite{Sznol16}. In order to design vehicle safety systems to prevent BTAR, a detailed understanding of the underlying injury mechanism is critical.
		
		In light of its severity and somewhat common occurrence, the mechanism of aortic injury in BTAR has been widely investigated, including such proposed mechanisms as stretching and torsion of the aorta \cite{Rindfleisch93}, a sudden rise in blood pressure \cite{Oppenheim18}, the water-hammer effect \cite{Lundevall64}, and the pinching of the aorta between the spine and anterior thorax \cite{Crass90}; the reader is referred to \citeA{Chapman01} and \citeA{Richens02} for an excellent discussion of these mechanisms and more. A more recent and promising mechanism, termed the dynamic self-pinch, has been proposed by \citeA{Lee07}. Their realistic numerical experiment simulated the conditions of BTAR caused by a frontal impact and accounted for the fluid-structure interactions on an in vitro model of the aorta. They identified a local transverse bend of the aorta wall produced near the isthmus region with sufficient shear to initiate an intimal tear and with locally elevated longitudinal stresses that can propagate the tear through the vessel wall. Although BTAR is also known to occur in lateral impacts, the dynamic self-pinch mechanism may also present itself in such a case.
		
		Interestingly, it has been debated whether abrupt deceleration alone (i.e.\ in the absence of chest trauma) is sufficient to cause rupture \cite{Chapman01}. Although, as \citeA{Lee07} point out, there is considerable evidence against this hypothesis, including Indy car crashes \cite{Melvin98} and cadaver sled tests \cite{Forman05} where accelerations in excess of $100g$'s were endured with no evidence of BTAR occurring. Nonetheless, although deceleration alone seems to be insufficient to cause rupture, this has no bearing on its actual importance in BTAR and the problem still deserves careful treatment. In fact, the effects of accelerations (or body forces) on fluid flow in general, particularly in the transverse direction as seen by the descending aorta during BTAR, is a fundamental problem in fluid mechanics that appears to have had very little attention. Furthermore, given the complexity of the geometry and motion of the aorta, the corresponding hemodynamics and fluid-structure interactions, and the multitude of impact scenarios a victim of BTAR may experience, it will prove useful to investigate the fundamental problem where the physical effects of the deceleration are isolated.
		
		\begin{figure}[!h]
			\centering
			\includegraphics[width = 0.96\textwidth]{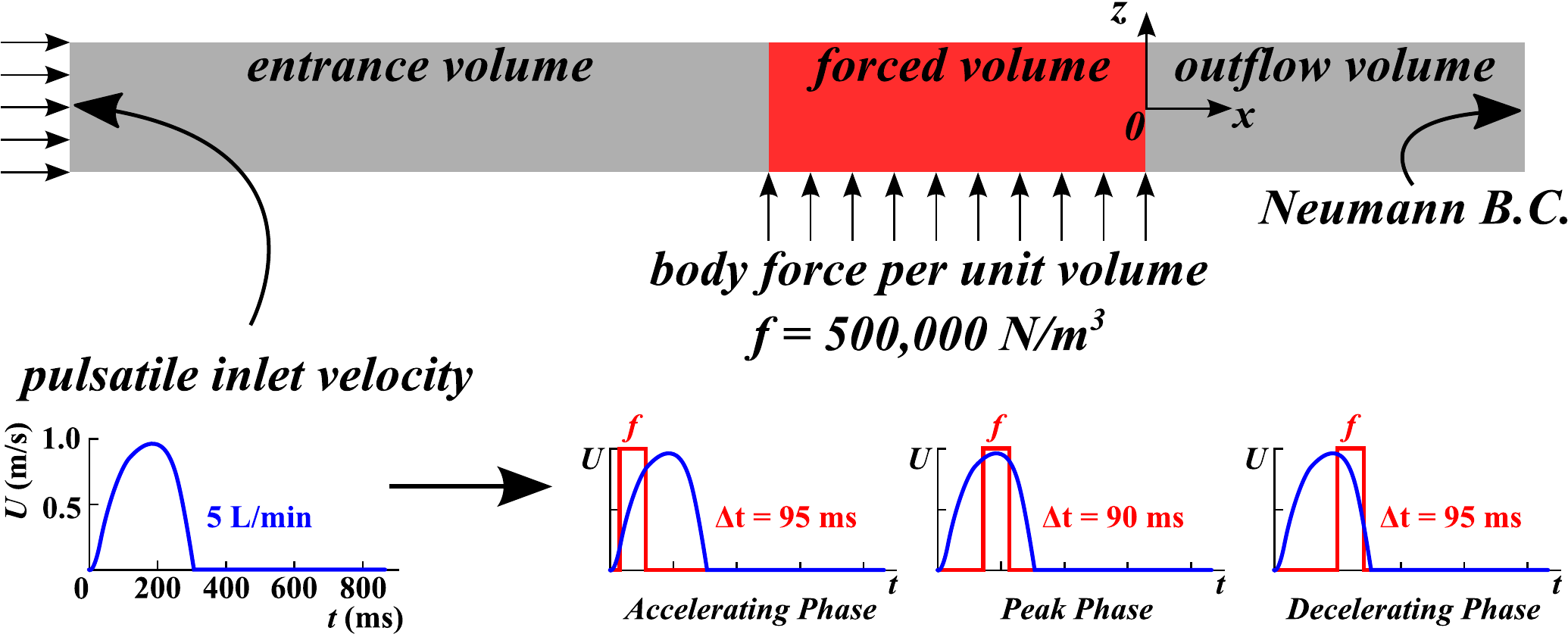}
		\caption{Schematic of the problem under study representing a circular pipe with a spatially uniform pulsatile inlet velocity excited by a localized impulsive body force applied on the red forced volume after the flow has fully developed. The body force is applied during three different instances of the pulsatile profile, namely during the accelerating phase, the peak phase, and the decelerating phase.}
		\label{fig:Schematic}
		\end{figure}
		
		Body forces acting on fluid flow in the transverse direction have been treated indirectly in the literature. The most popular example perhaps is flow behavior in pipes of single or multiple curvature, which has long been studied with cardiovascular and engineering flow applications. In such flows, the fluid experiences a centrifugal force as it flows around a bend and the velocity profile skews toward the outer wall. Secondary flows, such as the well-known Dean vortices \cite{Dean28}, also form in conjunction to this axial velocity skewing. While much attention has been given to these interesting flows, the body forces in these studies are time-independent. Very few studies consider time-dependent body forces, as is the case in BTAR. An interesting example is the experimental work of \citeA{Schilt96}, where a localized time-varying perturbation to the flow is investigated with applications to blood flow in coronary arteries. The study consisted of a sudden variation in the radius of curvature of a U-shaped circular pipe. Aside from the velocity profile skewing toward the outer wall, it was observed that the maximum skewing occurred approximately midway between the transitions from the minimum and maximum radii of curvature. In fact, they found that it is not the radius of curvature itself, but the dynamics of the curvature change that determined the skewing of the velocity profile. Also with applications to coronary flows, \citeA{Waters99} conducted a fundamental analytical investigation of curved elastic vessels with a pulsating curvature change in-phase with a pulsating flow. BTAR, of particular interest in this study, falls under a similar category of flows, namely blood flows subject to a time-varying localized acceleration.
		
		In this study, the effects of a transverse impulsive body force acting on a pulsatile flow in a straight circular pipe will be investigated. Although this problem will appear to be significantly simplified with regard to the physics of BTAR, the aim here is to address the need for an understanding of the physics of simple flows subject to transverse body forces. As is classically the case, the perspective gained from such fundamental studies are invaluable in obtaining a qualitative sense of the physics in more complex problems, such as the fluid dynamic contribution to BTAR. In what follows, the details of the fundamental physical problem and how it is treated will be described in section \ref{sec:Methods}. In particular, this work will investigate the effects of the body force applied during different instances of the pulsatile flow. The behavior of these flows in terms of their velocity field, kinetic energy, and coherent structures will be discussed in section \ref{sec:Res}. A summary will then be offered in the final section.
		
	\section{\label{sec:Methods}Methods}
		
		\subsection{\label{sec:ProbDef}Problem definition}
			
			In this fundamental numerical work, a spatially-uniform pulsatile velocity profile was supplied at the inlet of a straight, rigid pipe of circular cross-section with a diameter ($D$) of $0.022$ m (similar to that of a human aorta). The working fluid was modeled as incompressible and Newtonian with density ($\rho$) and dynamic viscosity ($\mu$) corresponding to those of blood, namely $1050$ kg/m$^3$ and $0.0035$ Pa$\cdot$s respectively \cite{Morris05}. The assumption of a Newtonian fluid is reasonable in large arteries due to the presence of larger shear rates (in excess of $100$ rad/s) which attenuate the non-Newtonian behavior of blood \cite{Fung81}. The rigid wall assumption for the aorta is also realistic as shown by \citeA{Jin13}. The pulsatile velocity waveform is the same as that used in \citeA{KeshavarzMotamed13} and has a peak Reynolds number ($\mathrm{Re}$) of $6358$ and an average $\mathrm{Re}$ of $4291$ over $0$ s $< t < 0.300$ s. Based on the peak $\mathrm{Re}$, an entrance length of $0.6$ m ($27.27D$) was determined using the classical entrance length equation for a turbulent flow \cite{Munson13} with some additional length added for safe-measure. On a strictly-bound volume of fluid where the flow has fully developed, a unidirectional rectangular-impulse body force per unit volume ($f$) of magnitude $500\, 000$ N/m$^3$ was applied. The body force magnitude and duration were selected to be within the range typically experienced in BTAR, namely $15$-$150$ $\rho g$ of force per unit volume acting for $50$-$100$ ms \cite{Chapman01}. The selection of a rectangular-impulse distribution to represent the body force acting on the thoracic aorta is used to simulate a worst case scenario where the entire volume in the forced section is subject to the totality of the body force, representing the limiting case of a generalized Gaussian-distribution-like force. Impacts were tested during the accelerating, peak, and decelerating phases of the pulsatile waveform with impact durations of $95$ ms ($0.040$ s to $0.135$ s), $90$ ms ($0.135$ s to $0.225$ s), and $95$ ms ($0.190$ s to $0.285$ s) respectively. See figure \ref{fig:Schematic} for a schematic of the problem (note the location of the coordinate system origin).
			
			The forced and outflow volume lengths were selected as $0.3$ m ($13.63D$) and a Neumann boundary condition was used on the velocity at the outlet of the pipe. The selection of this length for the forced volume was such that it would be in the range of common lengths of the thoracic aorta. The length of the outflow volume was selected to be short enough to ensure that the Neumann boundary condition has no effect on the upstream flow, yet long enough to capture the evolution of flow structures within the volume for the simulation time used. We define a dimensionless time variable $t^* = t/\Delta t$ such that, for a given impact duration $\Delta t$, it will acquire a value of $0$ at the moment of impact and $1$ at the end of the impact. The spatial coordinates are rendered dimensionless by normalization with the pipe diameter (i.e.\ $x^* = x/D$, $y^* = y/D$, and $z^* = z/D$). With this system, $x^*$ will have a value of $-13.63$ at the upstream boundary of the forced volume and $0$ at the downstream boundary. With the definitions of $x^*$ and $t^*$, the body force per unit volume may be expressed mathematically as:
			\begin{equation}
			\label{eq:body_force}
				f\left(N/m^3\right) = \left\lbrace
				\begin{array}{cc}
					500\, 000 & -13.63 < x^* < 0\quad \text{and} \quad 0 < t^* < 1 \\
					0 &  \text{otherwise}
				\end{array}\right. .
			\end{equation}
			
			As an analogy to the interaction parameter in magnetohydrodynamics, the interaction of the body force with the flow inertia is considered using a parameter $\Psi = fD/\rho U_0^2$, where $U_0$ is some characteristic velocity, such as the average inlet velocity over the impact duration. The values of $\Psi$ for the accelerating, peak, and decelerating cases are $22.5$, $11.7$, and $17.2$ respectively.
			
		\subsection{\label{sec:NumMeth}Numerical method}
		
		The governing equations in this study are the continuity and Navier-Stokes equations. Using the dimensionless parameters defined previously, the conservation equations can be non-dimensionalized as:
		\begin{gather}
			\nabla^* \cdot V^* = 0 \label{eq:continuity}\\
			\tau\frac{\partial{\mathbf{V}^*}}{\partial{t^*}} + (\mathbf{V}^* \cdot \nabla^*)\mathbf{V}^* = -\nabla^*P^* + \Psi\hat{\mathbf{f}} + \frac{1}{\mathrm{Re}}{\nabla^*}^2 \mathbf{V}^* \label{eq:Navier_Stokes}
		\end{gather}
		with
		$$
			\begin{array}{cccccc}
				\mathbf{V}^* = \displaystyle\frac{\mathbf{V}}{U_0} & \nabla^* = D\nabla & \tau = \displaystyle\frac{D/U_0}{\Delta{t}} & P^* = \displaystyle\frac{P}{\rho{U_0^2}} & \Psi = \displaystyle\frac{fD}{\rho{U_0^2}} & \mathrm{Re} = \displaystyle\frac{\rho{V}D}{\mu},
			\end{array}
		$$
		where the velocity vector is denoted by $\mathbf{V}$, pressure by $P$, and the unit body force vector by $\hat{\mathbf{f}}$ (this force term is of course only present during application of the body force). As the body force is a discontinuous function, the numerical solution of the governing equations benefits greatly from a finite volume approach \cite{LeVeque04}. The equations were solved using the commercial computational fluid dynamics software ANSYS FLUENT $12.1.4$ which uses the finite volume method. The simulations were performed using a transient pressure-based solver. Pressure and velocity were coupled using the PISO scheme. Gradients of fluid variables were interpolated on a least squares cell-based approximation. As the applied body force is known and rather large, pressure was interpolated with a body-force-weighted approach which has been used with success in convection problems involving substantial body forces \cite{Chandratilleke12, Rout12}. Momentum, turbulent kinetic energy, and specific dissipation rate were interpolated using the second-order upwind scheme. Time was discretized using a first-order implicit formulation.
		
		The inlet $\mathrm{Re}$ considered in this study ranges from $0$ to $6358$ and due to the magnitude and nature of the body force, a transition to turbulence is likely to occur. The problem therefore encompasses the laminar, transitional, and turbulent regimes. A notable turbulence model that has a reasonable capability of dealing with flows encompassing the three flow regimes is the $k$-$\omega$ model with shear stress transport \cite{MenterFerreira03}. This model was selected based on its stability and widespread use for general-purpose CFD, enhanced near-wall treatment (less restriction on $y^+$), and its capability of handling flows with adverse pressure gradients \cite{Menter94, MenterKuntz03, Versteeg07} which is expected to occur with application of the body force. This modeling approach has proven to be successful in many applications as discussed by \citeA{Esch03} and the obtained velocity field has previously been validated against experimental data for a variety of cardiovascular flows exhibiting a similar range of $\mathrm{Re}$ as in this study \cite{Ghalichi98, Ryval04, KeshavarzMotamed11, KeshavarzMotamed13}.
		
\begin{figure}[!h]
			\centering
			\includegraphics[width = 0.40\textwidth]{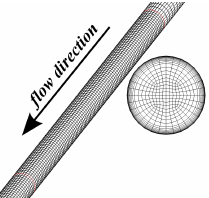}
		\caption{Schematic of the final mesh used for computation showing the longitudinal and cross-sectional mesh configuration. This mesh consists entirely of hexahedral elements with a total of $163\, 200$ elements. The face mesh contains $544$ elements and uses a growing boundary layer mesh of first layer thickness $5 \cdot 10^{-5}$ m (maximum corresponding $y^+ = 2.8$ between all three simulated impact conditions).}
		\label{fig:Mesh}
		\end{figure}

		In order to obtain adequate accuracy for a reasonable computation time, several different mesh configurations were tested. The selected mesh is similar to that used by \citeA{Lee07} and consisted entirely of hexahedral elements with a growing boundary layer mesh of first layer thickness $5 \cdot 10^{-5}$ m ($2.27 \cdot 10^{-3}D$) (see figure \ref{fig:Mesh}). The forced and outflow volumes were meshed twice as fine as the entrance volume. A grid independence study was then performed for a constant inlet $\mathrm{Re} = 5500$ (larger than the average $\mathrm{Re}$ of the pulsatile waveform) with a fine time step of $0.001$ s.
		
\begin{figure}[!t]
			\centering
			\includegraphics[width = 0.98\textwidth]{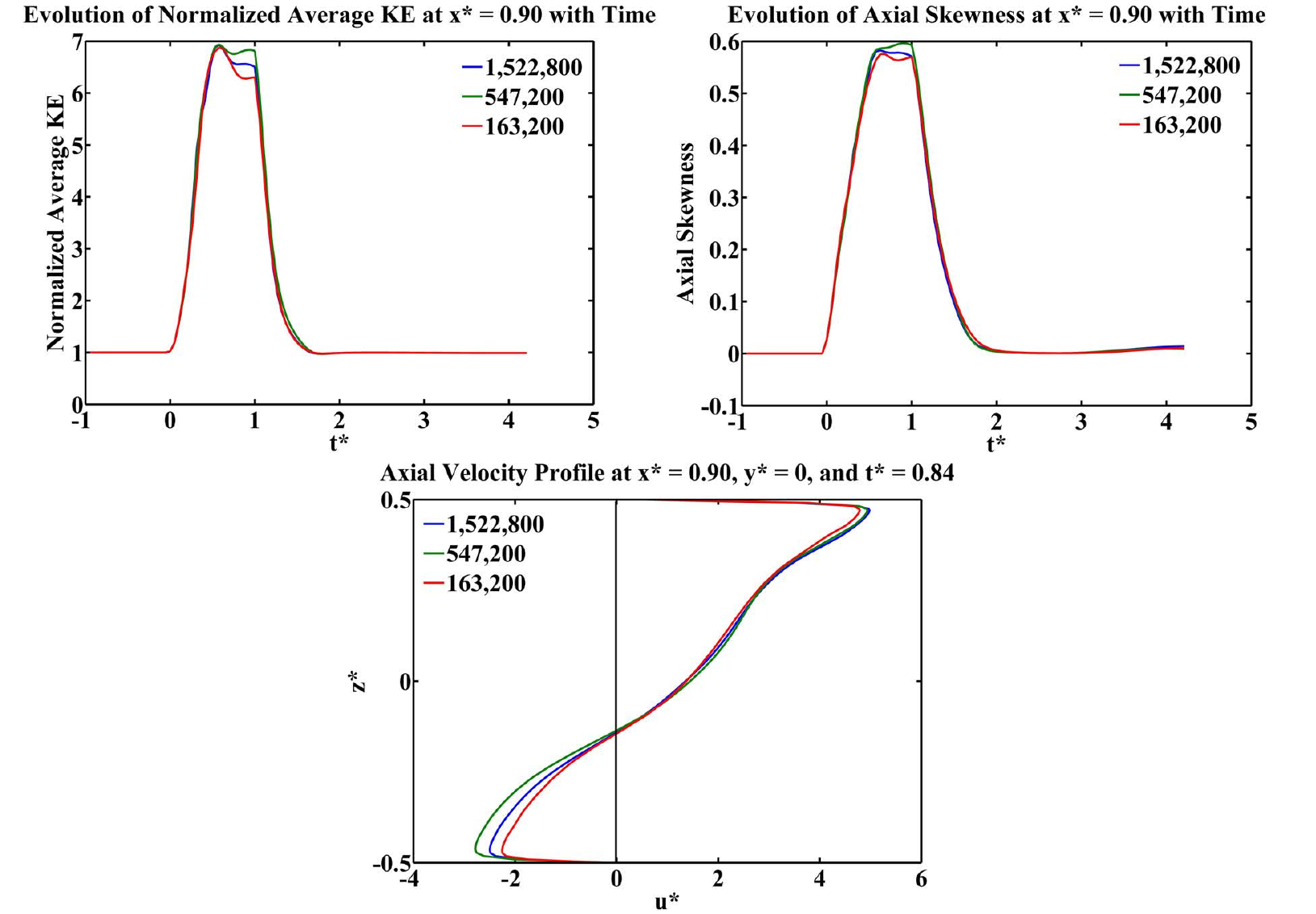}
			\caption{Normalized cross-sectionally averaged kinetic energy (top left) and axial skewness (top right) at $x^* = 0.90$ for a constant inlet $\mathrm{Re} = 5500$. During their decay, these quantities vary by less than $6$\% between the coarsest and finest meshes used in the grid convergence analysis. The largest errors observed for the normalized average kinetic energy were almost $10.5$\% occurring near the end of the impact ($t^* = 0.84$); the axial velocity profile (in the plane $y^* = 0$) was plotted at this instant of largest error, with a corresponding peak error of $4.7$\% between the coarsest and finest meshes.}
			\label{fig:Convergence}
		\end{figure}

		Three uniformly refined meshes were constructed with a refinement ratio of $1.4$; \citeA{Celik08} suggest $1.3$ or higher. The total number of elements amounted to $163\, 200$, $547\, 200$, and $1\, 522\, 800$. The evolution of cross-sectionally averaged kinetic energy and axial skewness (defined in section \hyperref[sec:Skewness]{\ref*{sec:Res}.\ref*{sec:Skewness}}) are of interest in this study and were examined at various cross-sections. During the decay of these quantities, less than $6$\% error was observed between the finest and the coarsest meshes with regard to the kinetic energy and axial skewness at the locations and times of interest, with errors in the velocity field being as much as $3$\%. The peak error observed for the cross-sectionally averaged kinetic energy was $10.5$\%, which occurred at $x^* = 0.90$ near the end of the impact ($t^* = 0.84$), with a corresponding maximum error in the velocity field of $4.7$\%. As this study is mainly interested in the decay of cross-sectionally averaged kinetic energy and axial skewness, and the velocity field has grid converged to within $5$\%, the numerical accuracy of the mesh with $163\, 200$ elements was judged sufficient and ultimately selected. Figure \ref{fig:Convergence} shows the results at one section of interest in this study.
		
		The convergence criteria for all flow properties were judged based on whether the absolute residuals for continuity, momentum components, turbulent kinetic energy, and specific dissipation rate all fell below $10^{-5}$. Time steps of $0.01$ s, $0.005$ s, and $0.001$ s were all tested for the selected mesh. No significant differences in the accuracy of the results were observed. For example, the velocity field differed by less than $0.03$\% between the finest and coarsest time steps before application of the body force. During and after application of the body force, up to $0.18$\% error was observed between the time steps of $0.001$ s and $0.005$ s. In light of these results, a variable time-step approach was applied. The time step was selected to be $0.01$ s for $t^* < 0$ (before the body force is applied) and reduced to $0.005$ s for $t^* \geq 0$ (after the body force is applied).
		
	\section{\label{sec:Res}Results}
		
		\subsection{\label{sec:Flows}Axial and secondary flows}
			
\begin{figure}[!b]
				\centering
				\subfloat[\label{fig:Pressure_Schematic}]{%
					\includegraphics[width=0.98\textwidth]{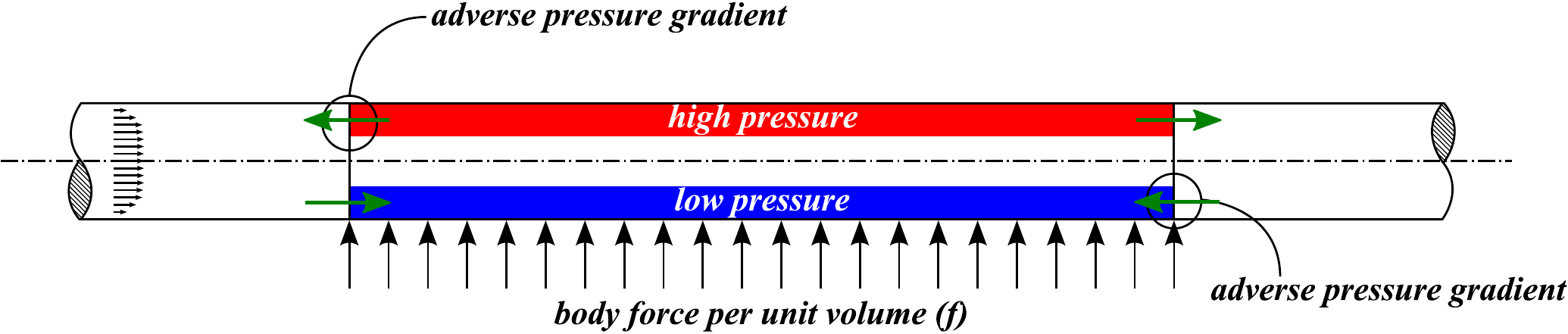}
				}\hfill
				\subfloat[\label{fig:Ax_2nd_Flows}]{%
					\includegraphics[width=0.98\textwidth]{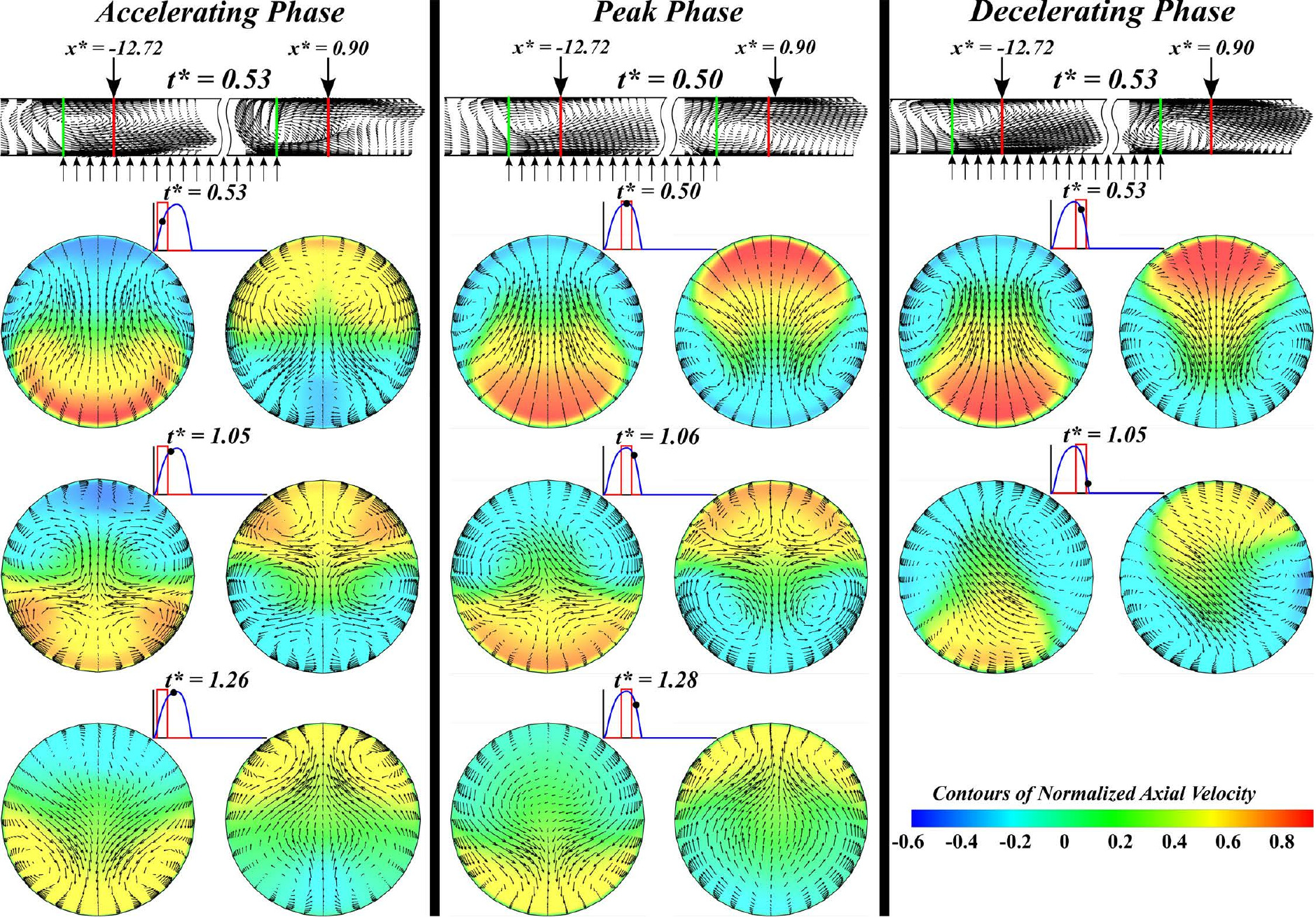}
				}
			\caption{(a) Schematic of the transverse pressure gradient induced by the action of the body force causing fluid to flow out of the forced volume at the top of the pipe and inward at the bottom. (b) Secondary flows and contours of normalized axial velocity at $t^* = 0.53$, $1.05$, and $1.26$ for the accelerating (left panel) and decelerating (right panel) cases and $t^* = 0.50$, $1.06$, and $1.28$ for the peak case (central panel). Two cross-stream counter-rotating vortices can be seen during the impact at $t^* \approx 0.5$ with secondary flows appearing as Dean-like vortices. After the impact ($t^* > 1$), more complex secondary flow structures develop to effectively restore the base flow configuration. It can also be seen that the time for the secondary flow to develop is longer from right to left (refer to the $t^* \approx 0.5$ sections for example).}
			\label{fig:Flow}
			\end{figure}

			At the moment the impulsive body force is applied, a transverse pressure gradient is developed in the forced volume, acting in the opposite direction of the force and equal in magnitude. The induced high pressure at the top of the pipe (positive $z^*$) generates a pressure gradient between the forced and unforced pipe sections, resulting in fluid being pushed out of the forced volume (see figure \ref{fig:Pressure_Schematic} and the online supplementary videos \href{http://stacks.iop.org/FDR/49/035510/mmedia}{\url{http://stacks.iop.org/FDR/49/035510/mmedia}}). The inverse effect, flow entering the forced volume, can be noticed at the bottom of the pipe (negative $z^*$). This results in the development of two cross-stream counter-rotating vortices in the vicinity of the boundaries of the forced volume (see the top panels in figure \ref{fig:Ax_2nd_Flows}). In fact, the effect of the body force is seen to be localized around the boundaries of the forced volume ($x^* = -13.63$ and $0$) while the flow in the center of the forced volume remains unperturbed for much of the impact duration. Figure \ref{fig:Ax_2nd_Flows} shows the two vortices for the three cases (i.e.\ the body force applied during the accelerating, peak, and decelerating phases of the pulsatile waveform) at approximately mid-impact ($t^* \approx 0.50$). In all cases, the vortices start at the wall and gradually move toward the centerline as they grow.
			
			It can be seen from the figure that the body force has a more substantial effect during the accelerating phase; stronger streamwise vortices develop at the boundaries of the forced volume section compared to the other two cases. This can be explained by the parameter $\Psi$, which is largest in the case of the accelerating flow due to the smaller average velocity over the impact duration, indicating that the body force dominates the fluid inertia more than in the other cases. The parameter $\Psi$ can also be thought of as a ratio of the transverse pressure gradient induced by the body force to the streamwise pressure gradient associated with the flow inertia. The larger the value of $\Psi$, the more the transverse pressure gradient dominates the streamwise pressure gradient.
			
			\begin{figure}[!b]
				\centering
				\includegraphics[width=0.99\textwidth]{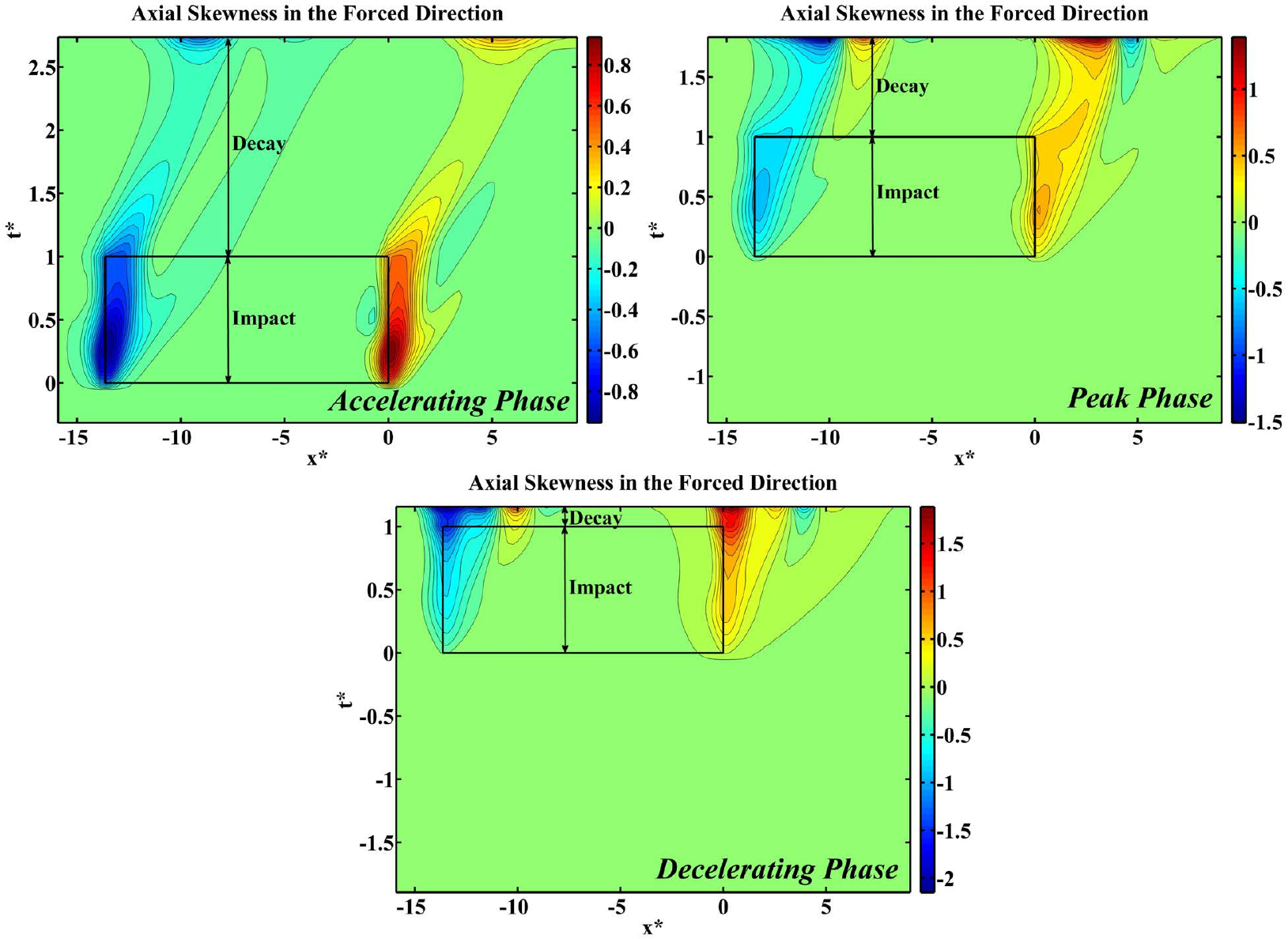}
			\caption{Contours of axial skewness for an impact occurring during the accelerating, peak, and decelerating phases. The maximum skewness is largest for the decelerating phase and occurs at the end of the impact, whereas it occurs before the end of the impact for the accelerating and peak phases.}
			\label{fig:Skewness}
			\end{figure}
			
			As the cross-stream vortices develop, secondary flows develop in the same regions simultaneously. Figure \ref{fig:Ax_2nd_Flows} shows the secondary flows with normalized axial velocity contours of two selected sections ($x^* = -12.72$ and $0.90$) at three selected times ($t^* \approx 0.53$, $1.05$, $1.26$) for each case (their temporal evolution can be seen in the online supplementary videos). For ease of viewing, the secondary flow vectors were normalized with respect to the maximum magnitude of the secondary flow for each given cross-section at each instant. For the upstream section ($x^* = -12.72$), high axial velocity occurs at the bottom of the pipe, where the low pressure draws the fluid inward, whereas it occurs at the top of the pipe at the downstream section ($x^* = 0.90$), where the high pressure forces the flow outward. During the impact ($0 < t^* < 1$), the secondary flows are Dean-like vortices which carry the fluid from high axial velocity regions to the near-wall region, effectively slowing it down to attain the unforced equilibrium conditions. After the impact ($t^* > 1$), more complex secondary flow structures develop to restore equilibrium. An example of the difference of the action of these two secondary flows can be seen by comparing the sections at $t^* = 1.06$ for the peak phase and $t^* = 0.53$ for the decelerating phase in figure \ref{fig:Ax_2nd_Flows}, depicting two similar instants in the flow with the former being after the application of the body force and the latter being during the application of the body force. Interestingly, the development of the secondary flow appears to be slowest for an impact occurring during flow acceleration and quickest for one occurring during flow deceleration. For lower $\mathrm{Re}$ flows (higher $\Psi$), the body force should have a more significant effect and one would therefore expect the secondary flow to develop much more quickly in the accelerating flow case since it has the lowest average impact $\mathrm{Re}$ of the three cases. In fact, this observation is in accordance with a study of ramp-type flows by \citeA{He00}. In their experimental work, a linearly increasing (as well as decreasing) flow rate is supplied through a circular pipe. They observed that the delay in radial turbulence propagation is larger for an accelerating flow than for a decelerating one. Since turbulence propagation occurs mainly by the diffusion of turbulent eddies, it is natural to expect a similar result for secondary flows such as those observed in this study.
			
		\subsection{\label{sec:Skewness}Axial skewness}
		
			The impulsive body force perturbs the velocity profile spatially and temporally. The skewness of the three-dimensional axial velocity profile for a given cross-section can be quantified in terms of a normalized first moment with respect to the direction of imposed skewing. This quantity was devised by \citeA{Griffith13} in their investigation of the effects of stenotic flows with eccentric stenosis. The axial skewness in the forced direction ($\mu_z$) is defined as:
			\begin{equation}
			\label{eq:Skewness}
				\mu_z = \displaystyle\frac{\displaystyle\int\int\frac{z}{D}\frac{u}{U}\mathrm{d}y\mathrm{d}z}{\displaystyle\int\int\frac{u}{U}\mathrm{d}y\mathrm{d}z} \quad \text{where} \quad U = \frac{1}{A}\int\int u\mathrm{d}y\mathrm{d}z
			\end{equation}
			
			Figure \ref{fig:Skewness} shows contours of axial skewness as a function of time and space for each case. From the contours, it can be seen that the effects of the body force are localized around the boundaries of the forced volume with skewness peaks occurring at these boundaries. When the body force is applied during the accelerating or peak phases of the pulsatile waveform, the maximum skewness for sections near the boundaries of the forced volume occurs prior to the end of the impact. The degree of influence of the body force (as characterized by $\Psi$ using the instantaneous inlet velocity magnitude) decreases for the accelerating flow whereas it is more or less constant for the impact occurring at the peak, which is why the peak skewness occurs earlier for the accelerating flow ($t^* = 0.26$) than for the peak flow ($t^* = 0.39$). When the body force is applied during the decelerating phase however, the peak skewness occurs precisely at the end of the impact ($t^* = 1$) as the degree of influence of the body force continuously increases for the decelerating flow. The development of the two cross-stream counter-rotating vortices and the secondary flows at these locations is the main contribution to the skewing. The propagation of these features downstream corresponds to propagation of the skewness downstream. When the body force ceases, the skewness continues to propagate downstream and decays with time, just as the streamwise vortices and secondary flows do. This is seen in the form of a decaying wave from the velocity field and in figure \ref{fig:Skewness} by a directed decay of the skewness contours. The sign of the skewness is positive for sections near the downstream boundary since high axial velocity exists at the top of the pipe (positive $z^*$), and negative near the upstream boundary due to the high axial velocity being at the bottom of the pipe (negative $z^*$).
			
			\begin{figure}[!t]
				\centering
				\includegraphics[width=0.91\textwidth]{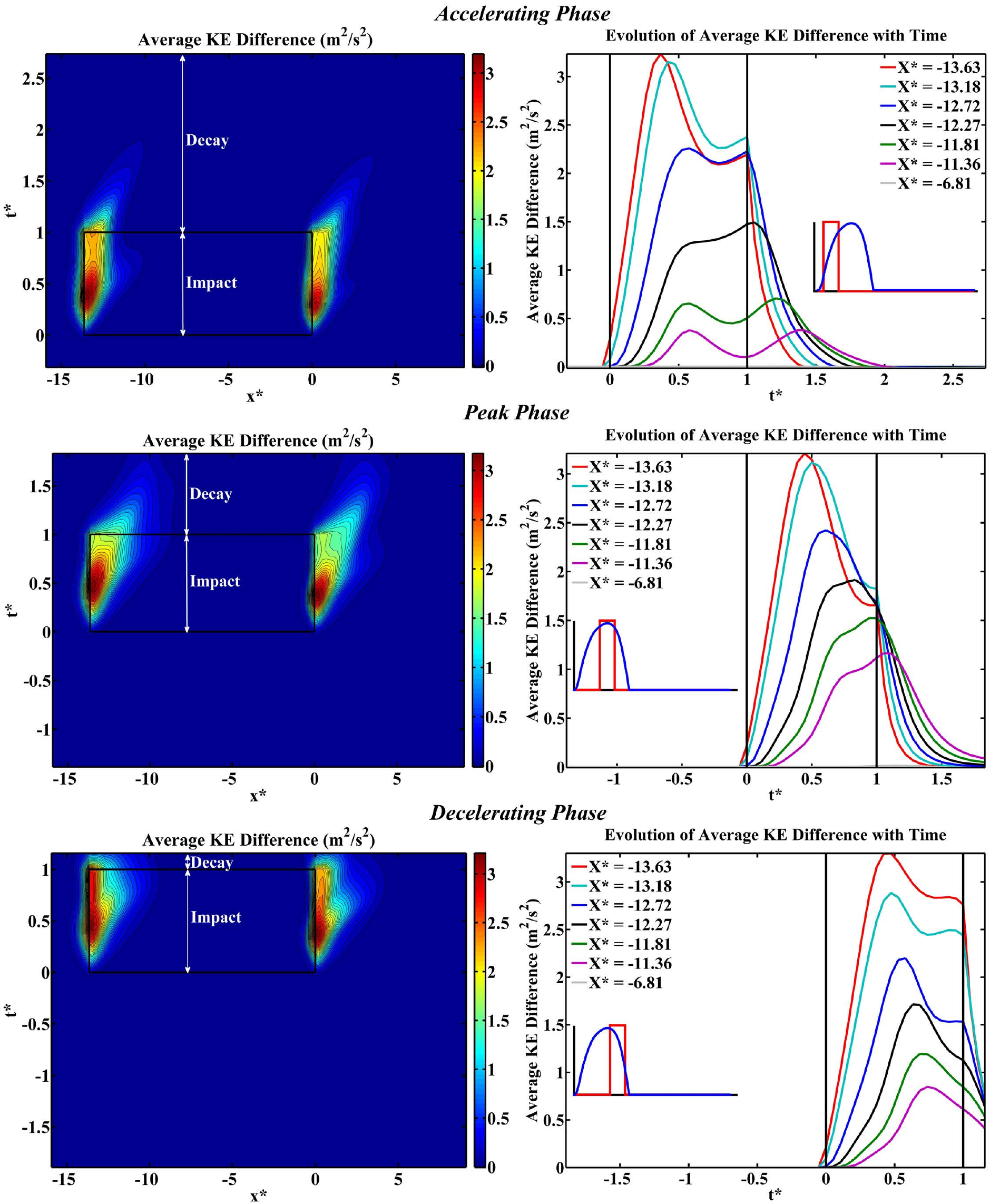}
			\caption{Contours of the difference between the cross-sectionally averaged kinetic energy (per unit mass) for an impact occurring during the accelerating, peak, and decelerating phases with the cross-sectionally averaged kinetic energy of the base flow. Plots at different upstream sections along the pipe are also shown. The kinetic energy is more localized spatially and elongated temporally for the accelerating flow whereas the opposite is seen for the decelerating flow.}
			\label{fig:KE}
			\end{figure}
			
		\subsection{\label{sec:KE}Cross-sectionally averaged kinetic energy}
		
			In figure \ref{fig:KE}, contour plots of cross-sectionally averaged kinetic energy (per unit mass) are shown for each case with the kinetic energy of the unforced flow subtracted off (denoted by ${\Delta}ke_{avg}$). The evolution of ${\Delta}ke_{avg}$ is also shown at six upstream sections ($x^* = -13.63$, $-13.18$, $-12.72$, $-12.27$, $-11.81$, $-11.36$, and $-6.81$); similar features are observed downstream. The maximum observed ${\Delta}ke_{avg}$ increase for the three cases exhibit similar magnitudes. For instance, at the upstream boundary of the forced volume, the maximum ${\Delta}ke_{avg}$ is $3.30$ m$^2$/s$^2$ (at $t^* = 0.37$ and $x^* = -13.41$), $3.28$ m$^2$/s$^2$ (at $t^* = 0.44$ and $x^* = -13.41$), and $3.30$ m$^2$/s$^2$ (at $t^* = 0.47$ and $x^* = -13.63$) for an impact occurring during the accelerating, peak, and decelerating phases respectively. The evolution of the kinetic energy at particular sections however, does not show a monotonic increase during the application of the body force but rather shows some variation. The decreases in kinetic energy appear to be associated with a transfer of energy from the axial (or mean) flow to the secondary flow whereas increases in kinetic energy appear to be associated with the shedding of energy from the secondary flow into the axial flow. Interestingly, at the upstream boundary, the maximum ${\Delta}ke_{avg}$ occurs at successively later times between each case, again in accordance with the degree of influence of the body force. The effects of the body force are still seen to be localized at the boundaries of the forced volume for all cases, however the evolution of kinetic energy is clearly different depending on the nature of the flow. For the accelerating flow, the ${\Delta}ke_{avg}$ contours appear more localized spatially and elongated temporally. The opposite is seen for the decelerating flow, where the contours are wider spatially and more localized temporally. The evolution of the ${\Delta}ke_{avg}$ at various sections in figure \ref{fig:KE} illustrates this spatial and temporal decay. Note that for $x^* = -6.81$ (the middle of the forced volume) ${\Delta}ke_{avg} \approx 0$, indicating that the effects of the body force have not propagated this far.
			
\begin{figure}[!h]
				\centering
				\subfloat[\label{fig:PL_A}]{%
					\includegraphics[width=0.48\textwidth]{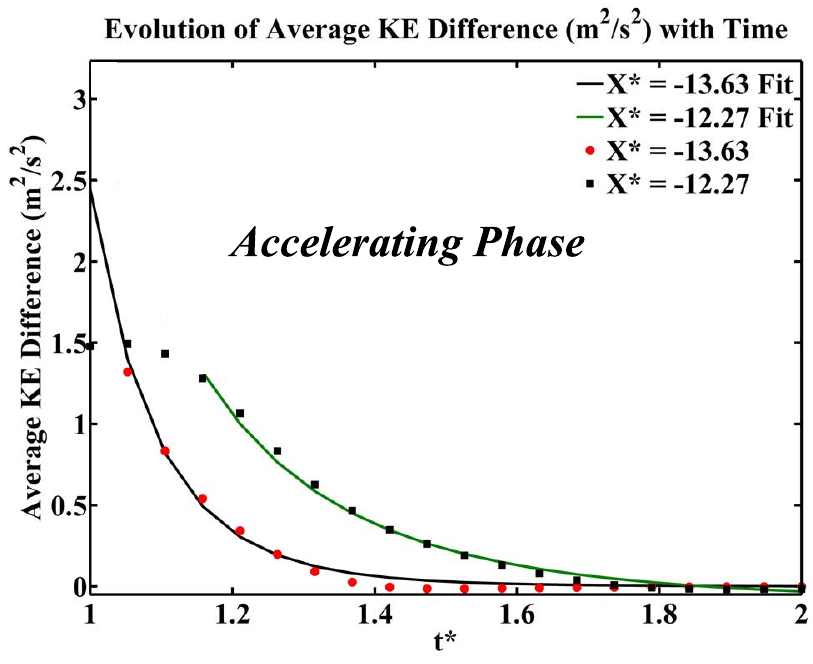}
				}\hfill
				\subfloat[\label{fig:PL_P}]{%
					\includegraphics[width=0.48\textwidth]{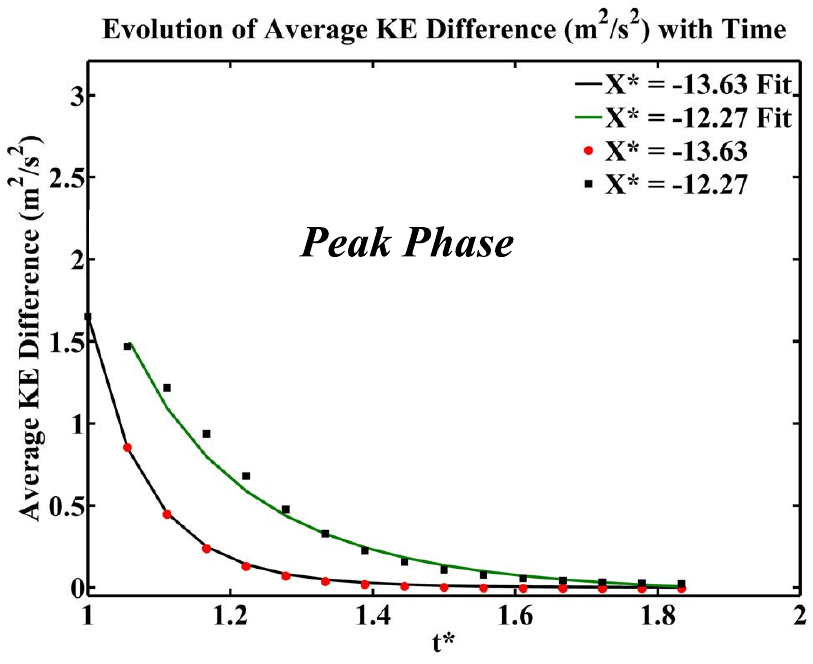}
				}
			\caption{Power law decay of kinetic energy (${\Delta}ke_{avg} \sim (t^*)^{-\alpha}$) at sections $x^* = -13.63$ and $-12.27$ for an impact occurring during (a) the accelerating phase ($\alpha = 10.8$ and $5.0$ respectively) and (b) the peak phase ($\alpha = 12.3$ and $5.6$ respectively) of the pulsatile inflow.}
			\label{fig:Power_Law}
			\end{figure}

			The temporal decay of the kinetic energy, as seen in figure \ref{fig:KE}, appears to follow a power law decay, namely ${\Delta}ke_{avg} \sim (t^*)^{-\alpha}$. This is illustrated in figure \ref{fig:Power_Law}, where such a power law fit is shown for the sections $x^* = -13.63$ and $-12.27$ after an impact occurring during the accelerating and peak phases; the power law fit could not be performed in the case of the impact occurring during the decelerating phase of the pulsatile inflow, since the flow was not given sufficient time to completely decay. The decay rates ($\alpha$) at $x^* = -13.63$ and $-12.27$ are respectively $10.8$ and $5.0$ for an impact during flow acceleration and $12.3$ and $5.6$ for one during peak inflow.
			
		\subsection{\label{sec:POD}Proper orthogonal decomposition of the velocity field}
			
			Proper orthogonal decomposition (POD) has become a powerful tool in the fluids community over the past $50$ years. In fluid dynamics, it is mainly used to decompose the velocity field into spatially orthogonal modes ranked by their energy content. The POD modes themselves provide a lower-order description of spatial coherence in the data, thereby significantly reducing the amount of data required for analysis of the dominant flow features. In order to obtain a deeper understanding of the effects of the body force, POD was applied on the velocity field for the three cases over a period including the impact and the first $100$ ms of the restoring flow after impact.
			
			\begin{figure}[!h]
				\centering
				\includegraphics[width=0.43\textwidth]{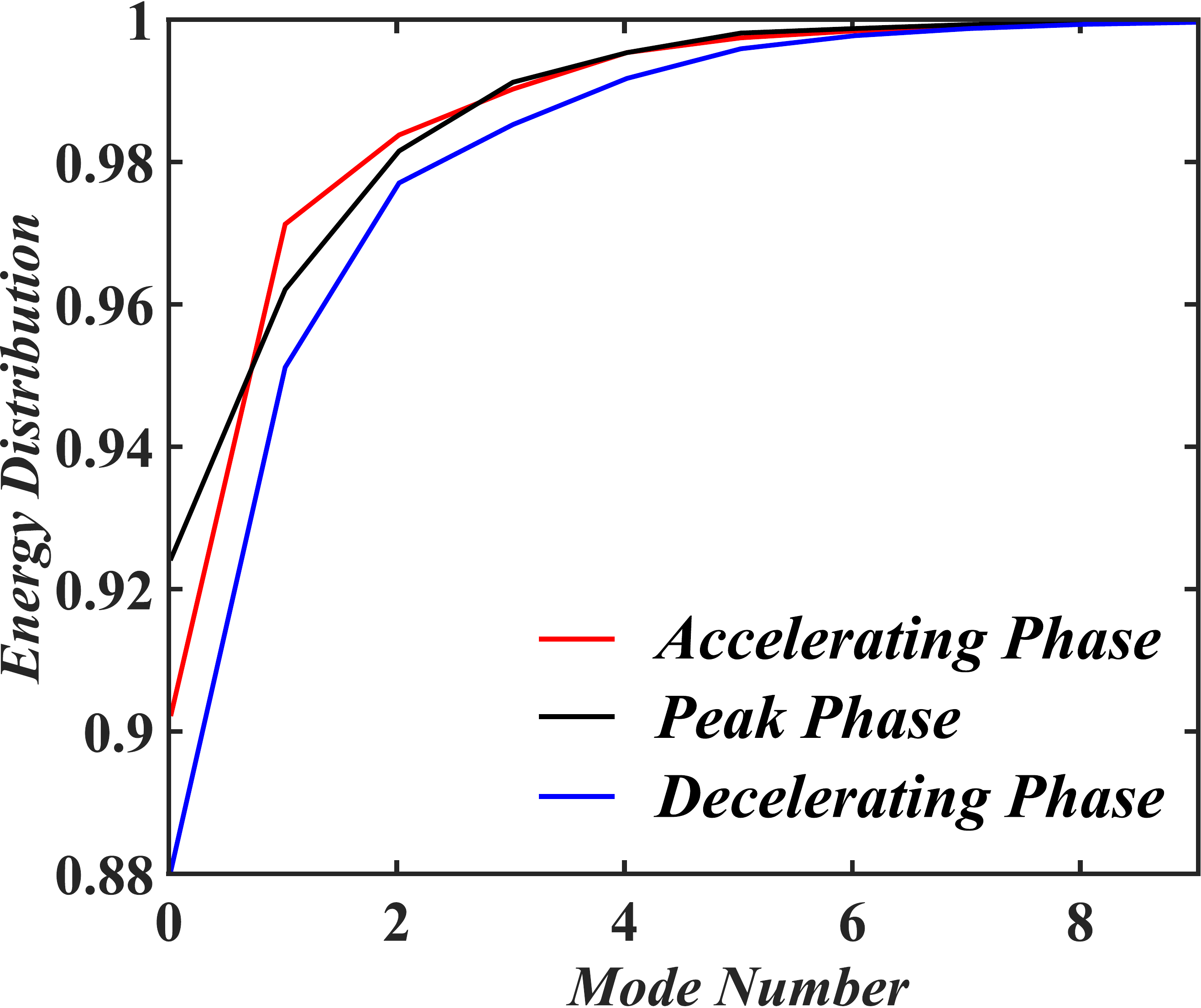}
			\caption{Plot of the cumulative fraction of total kinetic energy with proper orthogonal mode number for a body force applied during the accelerating (red), peak (black) and decelerating (blue) phases of the pulsatile inflow. The first three proper orthogonal modes capture $98.5$\%, $98.2$\%, and $97.8$\% of the total kinetic energy respectively for the accelerating, peak, decelerating cases.}
			\label{fig:POD_Energy}
			\end{figure}
			
\begin{figure}[!h]
				\centering
				\includegraphics[width=0.98\textwidth]{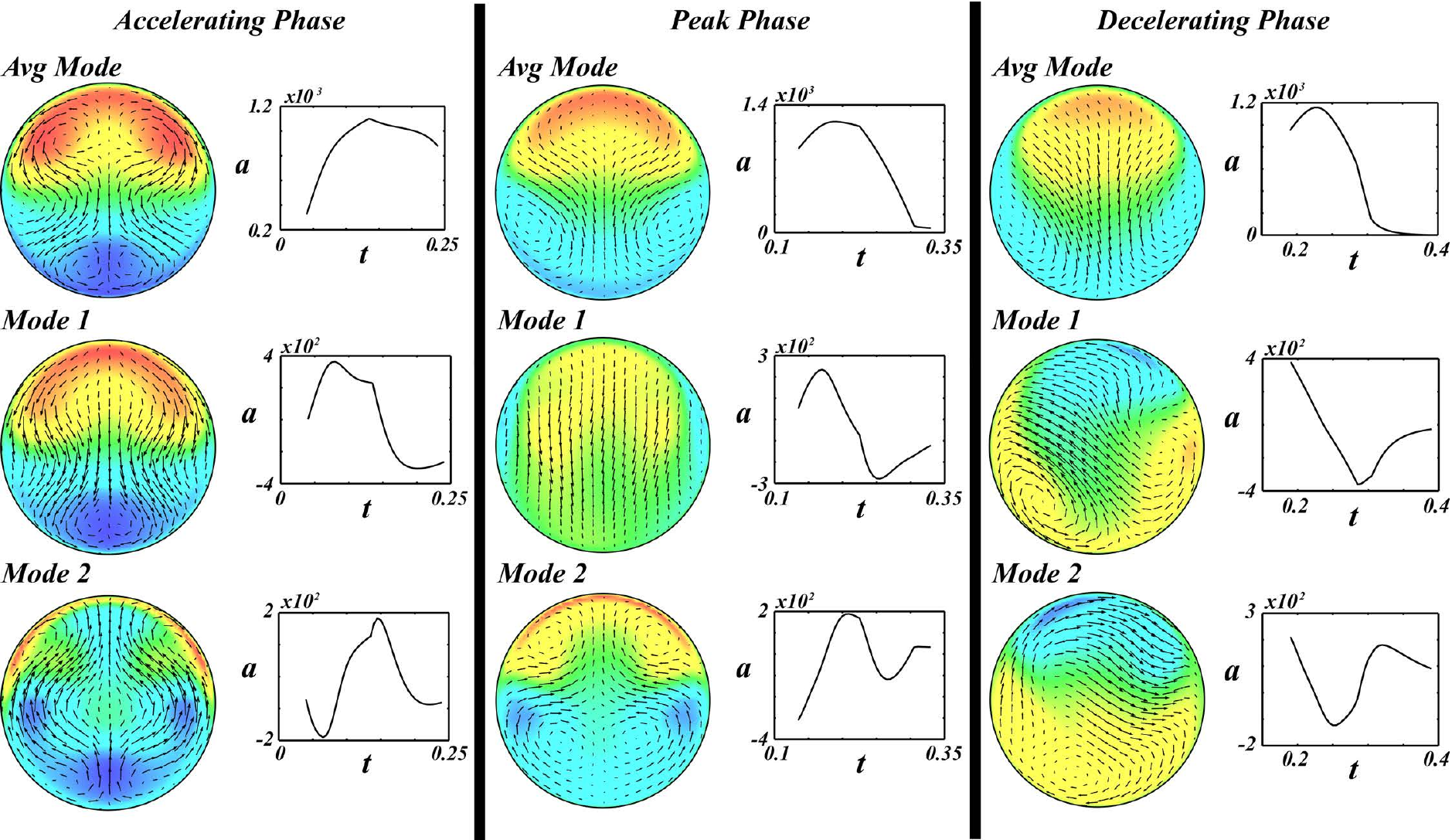}
			\caption{First three proper orthogonal modes for an impact applied during the accelerating, peak, and decelerating phases of the pulsatile inlet velocity profile. The coefficients `\textit{a}' represent the temporal evolution of its corresponding proper orthogonal mode.}
			\label{fig:POD_Modes}
			\end{figure}

			The cumulative fraction of total kinetic energy of the proper orthogonal modes for the three cases can be seen in figure \ref{fig:POD_Energy}. The first three proper orthogonal modes capture $98.5$\%, $98.2$\%, and $97.8$\% of the total kinetic energy per unit mass for the accelerating (Avg.\ Mode: $90.3$\%, Mode $1$: $6.9$\%, Mode $2$: $1.3$\%), peak (Avg.\ Mode: $92.4$\%, Mode $1$: $3.8$\%, Mode $2$: $2.0$\%), and decelerating (Avg.\ Mode: $88.1$\%, Mode $1$: $7.1$\%, Mode $2$: $2.6$\%) cases respectively. As such, only the first three modes will be investigated for the three cases. Furthermore, due to near anti-symmetric similarity between the two sections $x^* = -12.72$ and $x^* = 0.90$ displayed in figure \ref{fig:Flow}, only the latter section is considered here. Figure \ref{fig:POD_Modes} shows the first three modes at $x^* = 0.90$ for each case as well as the evolution of their amplitude in time; the demarcation between the impact and decay can be seen from the abrupt changes in the temporal evolution. For ease of viewing of the coherent structures, the proper orthogonal modes were interpolated onto a finer rectangular grid. The average mode for an impact occurring during the accelerating phase clearly shows two counter-rotating vortices acting on the high axial velocity regions of the flow and two weaker counter-rotating vortices acting on the regions with reversed axial flow. This equilibrium-restoring mode grows in amplitude until the end of the impact, after which it begins to decay as equilibrium is approached. The first and second modes demonstrate similar spatial structures and appear to counteract each other during the impact. These two modes carry the flow in opposite directions, with the second mode having vortices confined to the near-wall region effectively slowing the axial flow and the first mode carrying fluid directly from high axial velocity regions to reversed flow regions (the cross-stream vortex is clearly present in this case). Similar average mode behavior for an impact occurring during the peak and decelerating phases can be seen, though the spatial modes resemble simple Dean vortices and the high axial velocity regions are reduced. For all cases, once the body force is released, all modes can clearly be seen to tend toward zero as the unforced flow is gradually restored to its equilibrium state. In particular, for the peak and decelerating cases, a second abrupt change can be seen in the temporal evolution of the mode amplitudes. This is due to the pulsatile inlet velocity abruptly changing slope to remain identically zero.
			
	\section{\label{sec:Conc}Conclusion}
		
		This work presents a new class of fundamental problems in fluid dynamics that require further theoretical and experimental scrutiny and shows some applications in understanding BTAR. Pulsatile circular pipe flow subject to a localized transverse impulsive body force was investigated through numerical simulation for an impact occurring during the accelerating, peak, and decelerating phases of the pulsatile waveform. A parameter $\Psi$ was devised that quantifies the degree of influence of the body force by measuring the ratio of the induced transverse pressure gradient to the streamwise pressure gradient. Application of the body force induces counter-rotating cross-stream vortices located at the boundaries of the forced volume. Interesting secondary flow structures develop in conjunction to these cross-stream vortices. It was observed that the development of the secondary flow occurs later for an impact occurring during an accelerating flow and faster during a decelerating flow. Maximum skewing of the velocity profile was observed to occur during the impact for a body force applied during the accelerating and peak phases, while it occurred at the end of the impact for one applied during the decelerating phase. It was also observed to occur at successively later times for the accelerating, peak, and decelerating cases respectively. This can be understood in terms of the degree of influence of the body force which decreases successively for the respective cases. Proper orthogonal decomposition of the velocity field reveals dominant spatial structures acting to restore the flow to its unforced state. In relation to BTAR, this study demonstrates that complex secondary flows in the aorta may still develop without chest trauma in car accidents. Whether the corresponding transverse wall shear stress could actually cause rupture may not be the case, however it may have a significant role in explaining why rupture sometimes occurs away from the aortic isthmus such as in the ascending or descending aorta \cite{Chiesa03} or in the absence of any thoracic injury whatsoever \cite{Sevitt77}. With regard to the most damaging scenario, it is difficult to draw from this study due to the interplay between wall shear stress magnitude and duration. For instance, the most complex secondary flow structures were observed for an impact occurring during the accelerating phase whereas they developed earlier for one occurring during the decelerating phase.
	
	%
	%
	
	\section*{Supplemental Material}
	
		The reader is referred to \href{http://stacks.iop.org/FDR/49/035510/mmedia}{\url{http://stacks.iop.org/FDR/49/035510/mmedia}} for the supplemental material.
	
	%
	%
	%
	%
	
	\bibliographystyle{apacite}

\end{document}